\documentclass[iop]{emulateapj}
\usepackage{epsfig,amsfonts,amsmath,graphicx,natbib,apjfonts}
\def\ra#1#2#3{#1$^{\rm h}$#2$^{\rm m}$#3$^{\rm s}$}
\def\dec#1#2#3{$#1^\circ#2'#3''$}

\def\swift{{\it Swift}}
\def\xmm{{\it XMM-Newton}}
\def\chandra{{\it Chandra}}
\def\grb{GRB\,130603B}
\newcommand{\be}{\begin{equation}}
\newcommand{\ee}{\end{equation}}

\def\har{1}
\def\col{2}
\def\illa{3}
\def\illp{4}
\def\asu{5}
\def\ifa{6}
\def\pom{7}
\def\cas{8}

\shorttitle{Short GRB 130603B}
\shortauthors{Fong~et.~al.}

\begin{document}

\title{Short GRB\,130603B:  Discovery of a jet break in the optical and radio afterglows, and a mysterious late-time X-ray excess}

\author{ 
W.~Fong\altaffilmark{\har},
E.~Berger\altaffilmark{\har},
B.~D.~Metzger\altaffilmark{\col},
R.~Margutti\altaffilmark{\har},
R.~Chornock\altaffilmark{\har},
G.~Migliori\altaffilmark{\har},
R.~J.~Foley\altaffilmark{\illa}$^{,}$\altaffilmark{\illp},
B.~A.~Zauderer\altaffilmark{\har},
R.~Lunnan\altaffilmark{\har},
T.~Laskar\altaffilmark{\har},
S.~J.~Desch\altaffilmark{\asu},
K.~J.~Meech\altaffilmark{\ifa},
S.~Sonnett\altaffilmark{\ifa},
C.~Dickey\altaffilmark{\pom},
A.~Hedlund\altaffilmark{\pom},
P.~Harding\altaffilmark{\cas}
}

\altaffiltext{1}{Harvard-Smithsonian Center for Astrophysics, 60
Garden Street, Cambridge, MA 02138, USA}
\altaffiltext{2}{Department of Physics and Columbia Astrophysics Laboratory, Columbia University, New York, NY, 10027}
\altaffiltext{3}{Astronomy Department, University of Illinois at UrbanaChampaign, 1002 W. Green Street, Urbana, IL 61801 USA}
\altaffiltext{4}{Department of Physics, University of Illinois Urbana-Champaign, 1110 W. Green Street, Urbana, IL 61801 USA}
\altaffiltext{5}{School of Earth and Space Exploration, Arizona State University, P. O. Box 871404, Tempe, AZ, 85287-1404}
\altaffiltext{6}{Institute for Astronomy, University of Hawaii, 2680 Woodlawn Drive, Honolulu, HI 96822, USA}
\altaffiltext{7}{Pomona College, 610 N. College Ave., Claremont, CA, 91711}
\altaffiltext{8}{Department of Astronomy, Case Western Reserve University, Cleveland, OH 44106-7215, USA}

\begin{abstract}

We present radio, optical/NIR, and X-ray observations of the afterglow
of the short-duration \grb, and uncover a break in the radio and
optical bands at $\approx 0.5$~d after the burst, best explained as a
jet break with an inferred jet opening angle of $\approx
4-8^{\circ}$. GRB 130603B is only the third short GRB with a radio
afterglow detection to date, and the first time that a jet break is
evident in the radio band. We model the temporal evolution of the
spectral energy distribution to determine the burst explosion
properties and find an isotropic-equivalent kinetic energy of $\approx
(0.6-1.7) \times 10^{51}$~erg and a circumburst density of $\approx 5
\times 10^{-3}-30$~cm$^{-3}$. From the inferred opening angle of \grb,
we calculate beaming-corrected energies of $E_\gamma\approx
(0.5-2)\times 10^{49}$~erg and $E_K\approx (0.1-1.6)\times
10^{49}$~erg. Along with previous measurements and lower limits we
find a median opening angle of $\approx 10^{\circ}$. Using
the all-sky observed rate of $10$~Gpc$^{-3}$~yr$^{-1}$, this implies a
true short GRB rate of $\approx 20$~yr$^{-1}$ within 200~Mpc, the
Advanced LIGO/VIRGO sensitivity range for neutron star
binary mergers. Finally, we uncover evidence for significant excess
emission in the X-ray afterglow of \grb\ at $\gtrsim 1$~d and conclude
that the additional energy component could be due to fall-back
accretion or spin-down energy from a magnetar formed following the
merger.

\end{abstract}

\section{Introduction}

The broad-band afterglows of short-duration gamma-ray bursts (GRBs;
$T_{90}<2$ sec; \citealt{kmf+93}) provide a unique opportunity to study the
basic properties of these events: their energetics, circumburst
densities, and jet opening angles. While the energy scales and
densities provide fundamental insight to the explosion physics and
progenitors, the geometry provides direct information on the
population's true energy scale and event rate. The most likely
progenitors, neutron star-neutron star (NS-NS) and/or neutron
star-black hole (NS-BH) mergers \citep{elp+89,npp92,fbf10,ber10,fb13},
are the premier candidates of gravitational waves for Advanced
LIGO/VIRGO. Thus, an inference on the opening angle distribution
will aid our expectations for coincident electromagnetic-gravitational
wave detections.

Over the past $\sim 9$ years, the \swift\ satellite \citep{ggg+04} has
discovered $77$~short GRBs, with a median $\gamma$-ray fluence of
$\approx 2 \times 10^{-7}$~erg~cm$^{-2}$ ($15-150$~keV;
\citealt{gbb+08,nfp09}). Coupled with redshift measurements from their
host galaxies, this has demonstrated a range of isotropic-equivalent
$\gamma$-ray energies of $E_{\rm \gamma,iso} \approx
10^{48}-10^{52}$~erg \citep{ber07,nfp09}. Temporal monitoring of their
broad-band afterglows has led to a similarly broad range of inferred
isotropic-equivalent kinetic energies, $E_{\rm K,iso} \approx
10^{48}-10^{52}$~erg, circumburst densities of $\lesssim 1$~cm$^{-3}$
\citep{sbk+06,pan06,sdp+07,pmg+09,ber10,fbc+11}, and opening angles of
$\gtrsim 3-25^{\circ}$
\citep{ffp+05,gbp+06,sbk+06,fbm+12,mbf+12,nkg+12,bzl+13,sta+13}.

Collimation in GRBs is determined from temporal steepenings
in afterglow light curves, termed ``jet breaks'', which are expected
to be achromatic \citep{sph99,rho99,pan05}. Jet breaks in the light
curves of long GRBs translate to an opening angle distribution with a
range of $\theta_j\approx\!2-25^\circ$ and a median of $\approx 7^{\circ}$,
leading to beaming-corrected energies of $E_{\gamma}=[1-{\rm
cos}(\theta_j)]E_{\gamma,{\rm iso}} \approx 10^{51}$ erg
\citep{bfk03,fks+01,fb05,kb08,rlb+09}. For short GRBs, the search for
jet breaks has been more challenging, primarily due to the intrinsic
faintness of their afterglows, a direct reflection of their low energy
scales and circumburst densities. There are only three robust cases of
jet breaks for short GRBs based on well-sampled light curves thus far
(GRB\,051221A: \citealt{sbk+06}, GRB\,090426: \citealt{nkr+11};
GRB\,111020A: \citealt{fbm+12}), with $\theta_j \approx
3-8^{\circ}$. The non-detection of jet breaks to $\gtrsim 1$~day after
the burst has also provided lower limits on the jet opening angles of
$\gtrsim 3-25^{\circ}$
\citep{ffp+05,gbp+06,fbm+12,chp+12,mbf+12,bzl+13}, suggesting that
short GRBs may have wider jets than their long-duration counterparts.

In addition to an inference on the opening angles, afterglows can also
provide unique constraints on the energy scales and circumburst
densities through multi-wavelength detections and modeling of the
spectral energy distributions. However, of the $77$ \swift\ short GRBs
detected to present, only two have been detected in the radio band
(GRB\,050724A: \citealt{bpc+05}; GRB\,051221A: \citealt{sbk+06}). Both
of these events have inferred densities of $\approx 10^{-2}$~cm$^{-3}$
and isotropic-equivalent energies of $\approx 10^{51}$~erg. The
upgrade of the VLA with a ten-fold increase in sensitivity
\citep{pcb+11} provides a promising route to increased radio afterglow
detections, and thus substantially tighter constraints on these
properties.

Temporal afterglow information, particularly in the X-ray band, has
also revealed cases of anomalous behavior that do not fit with the
standard synchrotron picture of afterglows. For short GRBs, there have
been observed cases of flares at $\lesssim 1000$~s after the burst
\citep{gbp+06,mcg+11}, shallow decays attributed to energy injection
(e.g., \citealt{sbk+06}), putative early plateaus attributed to the
spin-down power of a hyper-massive and highly-magnetized neutron star
remnant \citep{rot+10,rom+13} and two cases of late-time X-ray
re-brightenings on $\sim $~day timescales (GRB\,050724A:
\citealt{gbp+06}; GRB\,080503: \citealt{pmg+09}).

Recently, the short-duration \grb\ sparked much interest, because its
bright optical afterglow enabled the first afterglow spectrum of a
short GRB and thus an unambiguous redshift determination of
$z=0.3565 \pm 0.0002$ \citep{cpp+13,dtr+13}. It also led to the first
claimed detection of a ``kilonova'' associated with a short GRB
\citep{bfc13,tlf+13}, providing direct evidence for a compact object
binary progenitor. Here, we present and model the broad-band afterglow
of \grb. In Section~2, we present the X-ray, optical/NIR and radio
data sets. In Section~3, we model the evolution of the spectral energy
distribution (SED) and constrain the burst explosion properties. In
Section~4, we discuss the steepening in the radio and optical light
curves, best explained as a jet break. In Section~5, we investigate
several possibilities to explain excess X-ray emission at $\gtrsim
1$~day. Finally in Section 6, we discuss \grb\ in the context of the
short GRB population, and investigate implications for the energy
scale and event rate.

Unless otherwise noted, all magnitudes in this paper are in the AB
system and are corrected for Galactic extinction in the direction of
the burst using $E(B-V)=0.02$ mag \citep{sfd98,sf11}. Reported
uncertainties correspond to $68\%$ confidence. We employ a standard
$\Lambda$CDM cosmology with $\Omega_M=0.27$, $\Omega_\Lambda=0.73$,
and $H_0=71$ km s$^{-1}$ Mpc$^{-1}$.

\section{Observations}

\tabletypesize{\scriptsize}
\begin{deluxetable*}{lccccccc}
\tablecolumns{8}
\tablewidth{0pc}
\tablecaption{GRB\,130603B Afterglow Photometry
\label{tab:obs}}
\tablehead {
\colhead {$\delta t$}          &
\colhead {Exposure Time}           &
\colhead {Telescope}           &
\colhead {Instrument}          &
\colhead {Band}               &
\colhead {$F_{\nu}$}      &
\colhead {$\sigma_\nu$}      &
\colhead {References}  \\
\colhead {(d)}                 &
\colhead {(hr)}                    &
\colhead {}                    &
\colhead {}                    &
\colhead {}                 &
\colhead {($\mu$Jy)}        &
\colhead {($\mu$Jy)}   &
\colhead {}                            
}
\startdata
\multicolumn{8}{c}{{\it X-rays}}  \\
\noalign{\smallskip}
\hline
\noalign{\smallskip}

$2.69$ & $5.14$ & \xmm\          & EPIC-pn & 1~keV & $2.44 \times 10^{-3}$ & $4.59 \times 10^{-4}$ & This work \\
$6.50$ & $8.38$ & \xmm\          & EPIC-pn & 1~keV & $8.25 \times 10^{-4}$ & $5.03 \times 10^{-4}$ & This work \\

\noalign{\smallskip}
\hline
\noalign{\smallskip}
\multicolumn{8}{c}{{\it Optical}} \\
\noalign{\smallskip}
\hline
\noalign{\smallskip}
$0.008$ & $0.05$ & Swift         & UVOT   & $V$    & $<199.5$             &                       & 1 \\
$0.089$ & $1.42$ & Swift         & UVOT   & $V$    & $<53.95$              &                       & 1 \\
$0.24$ & $0.50$ & NOT            & MOS    & $r$    & $12.59$               & $0.23$                & 2 \\
$0.25$ & $0.25$ & WHT            & ACAM   & $z$    & $25.35$               & $1.36$                & 1 \\
$0.27$ & $0.25$ & WHT            & ACAM   & $i$    & $16.44$               & $0.88$                & 1 \\
$0.28$ & $0.50$ & CAHA           & DLR-MKIII & $V$ & $8.32$                & $0.73$                & 1 \\
$0.29$ & $0.008$ & GTC           & OSIRIS & $r$    & $10.96$               & $0.20$                & 2 \\
$0.29$ & $0.25$ & WHT            & ACAM   & $g$    & $6.31$                & $0.34$                & 2 \\
$0.33$ & $0.40$ & Gemini-South   & GMOS   & $g$    & $5.30$                & $0.19$                & 3 \\
$0.34$ & $0.17$ & Magellan/Baade & IMACS  & $r$    & $8.64$                & $0.14$                & 4 \\
$0.37$ & $0.40$ & Gemini-South   & GMOS   & $i$    & $12.25$               & $1.18$                & 3 \\
$0.60$ & $0.14$ & Gemini-North   & GMOS   & $z$    & $6.54$                & $0.18$                & 2 \\
$0.60$ & $0.19$ & UKIRT          & WFCAM  & $K$    & $13.68$               & $1.32$                & 2 \\
$0.61$ & $0.14$ & Gemini-North   & GMOS   & $i$    & $4.53$                & $0.12$                & 2 \\
$0.61$ & $0.19$ & UKIRT          & WFCAM  & $J$    & $9.29$                & $1.12$                & 2 \\
$0.62$ & $0.14$ & Gemini-North   & GMOS   & $r$    & $2.88$                & $0.08$                & 2 \\
$0.62$ & $0.14$ & Gemini-North   & GMOS   & $g$    & $1.60$                & $0.06$                & 2\\
$1.30$ & $0.15$ & Gemini-South   & GMOS   & $r$    & $<0.30$               &                       & 3 \\
$1.30$ & $0.15$ & Gemini-South   & GMOS   & $i$    & $<0.58$               &                       & 3 \\
$1.34$ & $0.33$ & Magellan/Baade & IMACS  & $r$    & $<0.46$               &                       & 4 \\
$1.59$ & $0.17$ & Gemini-North   & GMOS   & $g$    & $<0.19$               &                       & 2 \\
$1.60$ & $0.17$ & Gemini-North   & GMOS   & $r$    & $0.21$                & $0.05$                & 2 \\
$1.61$ & $0.17$ & Gemini-North   & GMOS   & $i$    & $<0.48$               &                       & 2 \\
$1.61$ & $0.39$ & UKIRT          & WFCAM  & $J$    & $<1.25$               &                       & 2 \\
$1.62$ & $0.17$ & Gemini-North   & GMOS   & $z$    & $<1.00$               &                       & 2 \\
$2.32$ & $0.37$ & VLT            & HAWK-I & $J$    & $<1.25$               &                       & 2 \\
$3.26$ & $0.17$ & GTC            & OSIRIS & $r$    & $<0.33$               &                       & 2 \\
$4.26$ & $0.17$ & GTC            & OSIRIS & $r$    & $<0.23$               &                       & 2 \\
$7.30$ & $0.37$ & VLT            & HAWK-I & $J$    & $<1.45$               &                       & 2 \\
$8.23$ & $0.33$ & TNG            & DOLoRes & $r$   & $<1.15$               &                       & 1 \\
$8.25$ & $0.33$ & TNG            & DOLoRes & $i$   & $<0.52$               &                       & 1 \\
$8.41$ & $0.33$ & Magellan/Baade & IMACS  & $r$    & $<0.40$               &                       & This work \\
$9.37$ & $0.62$ & HST            & ACS    & F606W  & $<0.03$               &                       & 4 \\
$^{a}9.41$ & $0.73$ & HST        & WFC3   & F160W  & $0.17$                & $0.03$                & 2, 4 \\
$21.26$ & $0.67$ & TNG            & DOLoRes & $r$   & $<1.49$              &                       & 1 \\
$21.29$ & $0.67$ & TNG            & DOLoRes & $i$  & $<0.93$               &                       & 1 \\
$21.52$ & $0.28$ & MMT           & MMTCam & $r$    & $<1.91$               &                       & This work \\
$29.57$ & $0.73$ & HST           & WFC3   & F160W  & $<0.10$               &                       & 4 \\
$37.34$ & $0.75$ & Magellan/Clay & LDSS3  & $r$    & $<0.40$               &                       & This work \\
\hline
\noalign{\smallskip}
\multicolumn{8}{c}{{\it Radio}} \\
\noalign{\smallskip}
\hline
\noalign{\smallskip}

$0.37$ & $2.00$  & VLA           &        & $4.9$~GHz & $125.1$            & $14.4$                & This work \\
$0.37$ & $2.00$  & VLA           &        & $6.7$~GHz & $118.6$            & $9.1$                 & This work \\
$1.43$ & $1.00$  & VLA           &        & $4.9$~GHz & $<56.7$            &                       & This work \\
$1.43$ & $1.00$  & VLA           &        & $6.7$~GHz & $64.9$             & $15.2$                & This work \\
$1.44$ & $1.00$  & VLA           &        & $21.8$~GHz & $<50.0$           &                       & This work \\
$4.32$ & $2.00$  & VLA           &        & $4.9$~GHz & $<51.0$            &                       & This work \\
$4.32$ & $2.00$  & VLA           &        & $6.7$~GHz & $<25.8$            &                       & This work \\
$84.31$ & $1.00$ & VLA           &        & $4.9$~GHz & $<68.6$            &                       & This work \\
$84.31$ & $1.00$ & VLA           &        & $6.7$~GHz & $<33.7$            &                       & This work 

\enddata
\tablecomments{All upper limits correspond to $3\sigma$
confidence. Optical flux densities are corrected for Galactic
extinction in the direction of the burst, but are not corrected
for extinction in the rest-frame of the burst. \\
$^{a}$ The reported flux and uncertainty are of the claimed kilonova detection \citep{bfc13,tlf+13}. \\
{\bf References:} (1) \citealt{dtr+13}; (2) \citealt{tlf+13}; (3) \citealt{cpp+13}; (4) \citealt{bfc13} }
\end{deluxetable*}

\grb\ was detected on 2013 June 3 at 15:49:14 UT by the Burst Alert
Telescope (BAT; \citealt{bbc+05}) on-board the \swift\ satellite
\citep{gcn14735}, and by {\it Konus-Wind} \citep{gcn14771}. \swift/BAT
localized the burst to RA=\ra{11}{28}{53.2} and
Dec=$+$\dec{17}{03}{48.2} (J2000) with $1.0'$ accuracy ($90\%$
containment; \citealt{gcn14741}). The $\gamma$-ray emission consists
of a single pulse with a duration of $T_{90}=0.18 \pm 0.02$~s
($15-150$~keV; \citealt{gcn14741}), a fluence of $f_{\gamma}=(6.6 \pm
0.7) \times 10^{-6}$~erg~cm$^{-2}$ ($20-10^{4}$~keV;
\citealt{gcn14771}), and a peak energy of $E_{\rm pk} = 660 \pm
100$~keV \citep{gcn14771}. The spectral lags are $0.6 \pm 0.7$~ms
($15-25$~keV to $50-100$~keV) and $-2.5 \pm 0.7$~ms ($25-50$~keV to
$100-350$~keV; \citealt{gcn14746}). The combination of the duration,
high peak energy, and zero (or negative) lag unambiguously classifies
\grb\ as a short-hard burst. At $z=0.3565$ \citep{cpp+13,dtr+13}, the
isotropic-equivalent gamma-ray energy is $E_{\gamma,{\rm iso}}\approx
2.1 \times 10^{51}$~erg ($20-10^{4}$~keV, rest-frame).

\subsection{X-rays}
\label{sec:xray}

Observations with the X-ray Telescope (XRT; \citealt{bhn+05}) on-board
\swift\ commenced at $\delta t=59$ s (where $\delta t$ is the time
after the BAT trigger) and revealed a fading, uncatalogued X-ray
source with a UVOT-enhanced position of RA=\ra{11}{28}{48.15} and
Dec=$+$\dec{17}{04}{16.9} (J2000) and an uncertainty of $1.4''$ radius
($90\%$ containment; \citealt{gtb+07,ebp+07,ebp+09}). The source faded
below the XRT detection threshold by $\delta t \approx 2$~d. We
analyze the XRT data using the HEASOFT package (v.6.13) and relevant
calibration files. To generate a count-rate light curve, we apply
standard filtering and screening criteria (see \citealt{mzb+13}),
ensuring a minimum signal-to-noise ratio of $4$ for each temporal bin.

We initiated a target-of-opportunity program on \xmm\ (ID: 072257, PI:
Fong) with the European Photon Imaging Camera (EPIC-PN) and obtained
observations at $\delta t \approx 2.7$~d and $\approx 6.5$~d with net
exposure times of $18.5$~ks and $30.2$~ks, respectively. We analyze
these data using standard routines in the Scientific Analysis System
(SAS) and detect a fading source coincident with the XRT position. We
find $\approx 60$ counts ($\sim 4\sigma$) in the first observation in
a $30''$-radius aperture, and $\approx 10$ counts ($3\sigma$) in the
second observation in a $15''$-radius aperture, where the aperture
size is adjusted to maximize the signal-to-noise ratio.

To determine the flux calibration, we model the XRT data with an
absorbed single power-law spectrum, using the Galactic neutral
Hydrogen absorption column ($N_{\rm H,MW}=1.93\times
10^{20}\,\rm{cm^{-2}}$; \citealt{kbh+05}). We fit for the photon
index, $\Gamma$ and the intrinsic Hydrogen absorption column ($N_{\rm
H,int}$ at $z=0.3565$). Using all of the available XRT data, we find
$N_{\rm{H,int}}=(2.4 \pm 0.4) \times 10^{21}\,\rm{cm^{-2}}$ and
$\Gamma=2.2 \pm 0.1$. We apply these spectral parameters to the XRT
and XMM data. The resulting fluxes from XMM are listed in
Table~\ref{tab:obs}.

\subsection{Optical}

Subsequent to the discovery of the X-ray afterglow \citep{gcn14735},
ground-based optical/NIR observations began at $\delta t \approx
2.7$~hr to search for an optical counterpart
\citep{bfc13,cpp+13,dtr+13,tlf+13}. We initiated four sets of $r$-band
observations of \grb\ using instruments on the twin Magellan 6.5-m
telescopes, spanning $\delta t \approx 8.1$~hr to $\approx 37$~d
(Table~\ref{tab:obs}). The description of these observations and the
optical afterglow discovery are provided in \citet{bfc13}. In addition, we
obtained a set of $r$-band observations with MMTCam on the $6.5$-m
Multi-Mirror Telescope (MMT) and processed these data using standard
procedures in IRAF. All of the published optical/NIR afterglow
photometry \citep{bfc13,cpp+13,dtr+13,tlf+13}, are listed in
Table~\ref{tab:obs}.

\begin{figure*}
\begin{minipage}[c]{\textwidth}
\tabcolsep0.0in
\includegraphics*[width=0.5\textwidth,clip=]{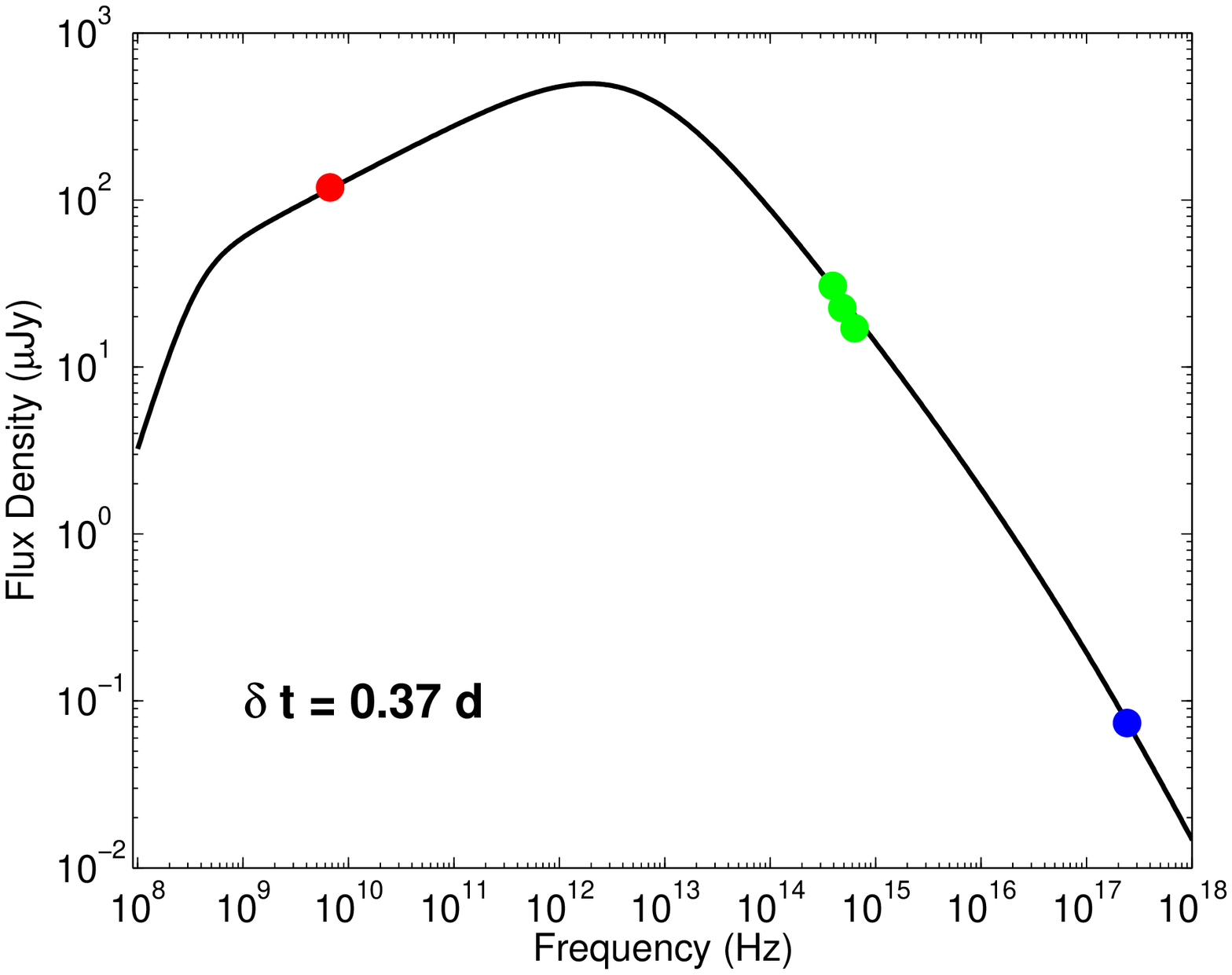} 
\includegraphics*[width=0.5\textwidth,clip=]{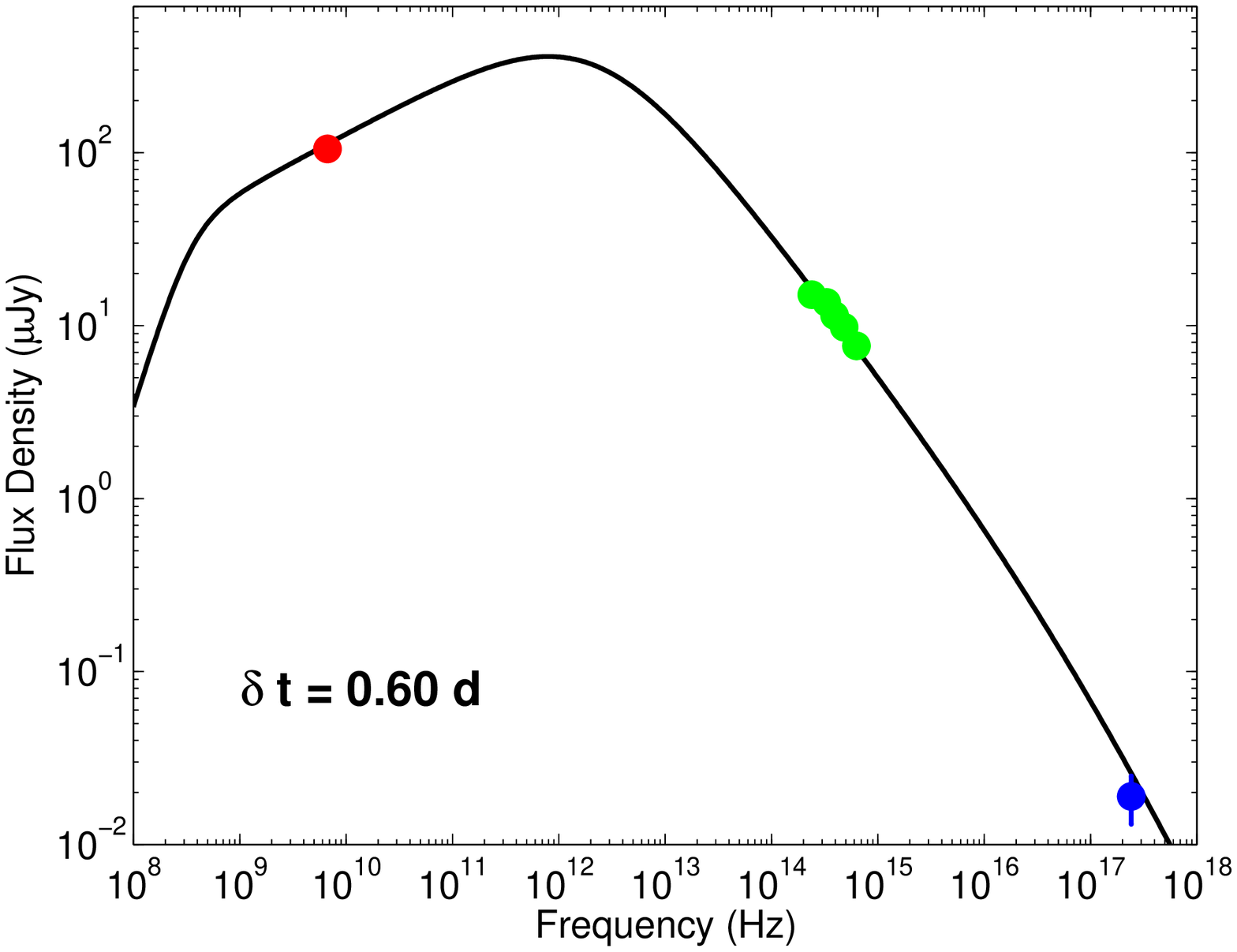} 
\end{minipage}
\caption{Radio (red; 6.7~GHz), optical/NIR (green; $griJ$-band;
\citealt{bfc13,cpp+13,tlf+13}) and X-ray (blue; 1~keV) afterglow
spectral energy distributions of \grb\ at $\delta t = 0.37$~d (left)
and $0.60$~d (right). The $grizJ$ afterglow observations are
corrected for $A_V^{\rm host}=1$~mag. Error bars correspond to
$1\sigma$ confidence. A representative best-fit model (black line) is
shown in each panel.
\label{fig:sed}}
\end{figure*}

\subsection{Radio}

We observed the position of \grb\ with the Karl G. Jansky Very Large
Array (VLA; Program 13A-046, PI: Berger) starting at $\delta t \approx
8.8$~hr at a mean frequency of $5.8$ GHz (upper and lower side-bands
centered at $6.7$~GHz and $4.9$~GHz) using 3C286 and J1120-1420 for
bandpass/flux and gain calibration, respectively. We follow standard
procedures in the Astronomical Image Processing System (AIPS;
\citealt{gre03}) for data calibration and analysis. We detect a source
located at RA=\ra{11}{28}{48.15}, Dec=$+$\dec{17}{04}{18.0} (J2000;
$\delta$RA$=0.21''$, $\delta$Dec$=0.14''$), consistent with the
optical and X-ray afterglow positions. We obtained three subsequent
$5.8$~GHz observations at $\delta t \approx 1.4$~d, $\approx 4.3$~d,
and $\approx 84.3$~d in which the source faded, indicating that it is
the radio afterglow. We also obtained observations at a mean frequency
of $21.8$~Hz at $\delta t \approx 1.4$~d (using J1118+1234 as the
phase calibrator), where the afterglow is not detected.

We measure flux densities for the upper and lower side-bands
from each epoch using AIPS/{\tt JMFIT}, and calculate $3\sigma$ upper
limits from source-free regions using AIPS/{\tt IMSTAT}. The radio
afterglow detections and upper limits are listed in
Table~\ref{tab:obs}.

\section{Broad-band Afterglow Properties}
\label{sec:sed}

We utilize the broad-band afterglow observations to constrain the
explosion properties and circumburst environment of \grb. We emphasize
that this is only the third radio afterglow detection of a short GRB,
thereby enabling substantially tighter constraints on the physical
properties than for the majority of short GRBs. We adopt the standard
synchrotron model for a relativistic blastwave in a constant density
medium (ISM; \citealt{sph99,gs02}). This model provides a mapping from
the broad-band afterglow flux densities to physical parameters of the
explosion and circumburst environment: isotropic-equivalent kinetic
energy ($E_{\rm K,iso}$), circumburst density ($n_0$), fractions of
post-shock energy in radiating electrons ($\epsilon_e$) and magnetic
fields ($\epsilon_B$), and the electron power-law distribution index
($p$), with $N(\gamma)\propto \gamma^{-p}$ for $\gamma \gtrsim
\gamma_{\rm min}$.

To determine the locations of the break frequencies with respect to
the X-ray, optical and radio bands, we first compare the
spectral indices from the afterglow observations. From the X-ray
spectral fit, we find $\beta_{X} \equiv 1-\Gamma = -1.2 \pm 0.1$
(Section~\ref{sec:xray}), while from the optical $griz$-band afterglow
photometry at $\delta t \approx 0.6$~d (Table~\ref{tab:obs};
\citealt{cpp+13,tlf+13}), we measure an observed spectral index of
$\beta_{\rm opt, obs} = -2.0 \pm 0.1$. The NIR spectral index measured
from the $JK$-bands at the same epoch is substantially shallower,
$\beta_{\rm NIR} = -0.6 \pm 0.2$, indicating that the optical flux is
suppressed by extinction in the host galaxy. To
determine the amount of extinction, we apply a Milky Way extinction
curve \citep{ccm89} to the $grizJK$ photometry, fitting for the
spectral index and rest-frame extinction ($A_V^{\rm host}$), and find
best-fit values of $\beta_{\rm opt} = -0.84 \pm 0.10$ and $A_{V}^{\rm
host} = 1.0 \pm 0.1$~mag, similar to the values found from other
analyses \citep{dtr+13,jxf+13}. Using standard Galactic
relations between the intrinsic hydrogen column density and rest-frame
extinction \citep{ps95,wat11}, we find an inferred extinction of
$A_V^{\rm host} = 0.9-1.6$~mag from the $N_{\rm H,int}$ value
determined in Section~\ref{sec:xray}, consistent with $A_{V}^{\rm
host} \approx 1$~mag. A comparison of $\beta_{\rm opt}$ and $\beta_{X}$
indicates that the cooling frequency, $\nu_c$, lies between the
optical and X-ray bands, and that $p = 2.55 \pm 0.15$.

To determine the location of the self-absorption and peak frequencies
($\nu_a$ and $\nu_m$, respectively) with respect to the optical and
radio bands, we compare the radio spectral slope determined from VLA
observations at $\delta t \approx 0.37$~d and $1.43$~d
(Table~\ref{tab:obs}) to $\beta_{\rm opt}$. We find that $\beta_{\rm
rad}$ does not match the optical slope and furthermore is consistent
with $F_\nu \propto \nu^{1/3}$, as expected for $\nu_{a} < \nu_{\rm
rad} < \nu_m$.

Knowing the relative locations of the break frequencies with respect
to our observing bands, we use the standard synchrotron model to
determine allowable ranges for the physical parameters. We use the
afterglow SED at two common epochs, $\delta t = 0.37$~d and $0.60$~d,
where the optical/NIR fluxes are corrected for $A_V^{\rm host}=1$~mag
(Figure~\ref{fig:sed}). We also use the constraints
$\epsilon_e,\epsilon_B<1/3$ and determine:

\begin{mathletters}
\begin{eqnarray}
3.3 \times 10^{8} < \nu_{a} < 2.6 \times 10^{9} \text{ Hz}
\label{eqn:constr1} \\
5.8 \times 10^{-2} < \epsilon_e < 1/3 
\label{eqn:constr2} \\
2.0 \times 10^{-3} < \epsilon_B < 1/3 
\label{eqn:constr3} \\
0.6 \times 10^{51} \text{ erg} < E_{\rm K,iso} < 1.7 \times 10^{51} \text{ erg}
\label{eqn:constr4} \\
4.9 \times 10^{-3} \text{ cm}^{-3} < n_0 < 30 \text{ cm}^{-3}
\label{eqn:constr5}
\end{eqnarray}
\end{mathletters}

\noindent where the ranges are set by the uncertainty in $\nu_a$ and
the conditions that $\epsilon_e,\epsilon_B<1/3$. We note that the inferred
isotropic-equivalent kinetic energy is comparable to $E_{\gamma,{\rm
iso}} \approx 2.1 \times 10^{51}$~erg. We show representative model
SEDs along with the broad-band afterglow observations in
Figure~\ref{fig:sed}.

\section{Afterglow Evolution: Jet Break}
\label{sec:jb}

\begin{figure*}
\includegraphics*[width=0.5\textwidth,clip=]{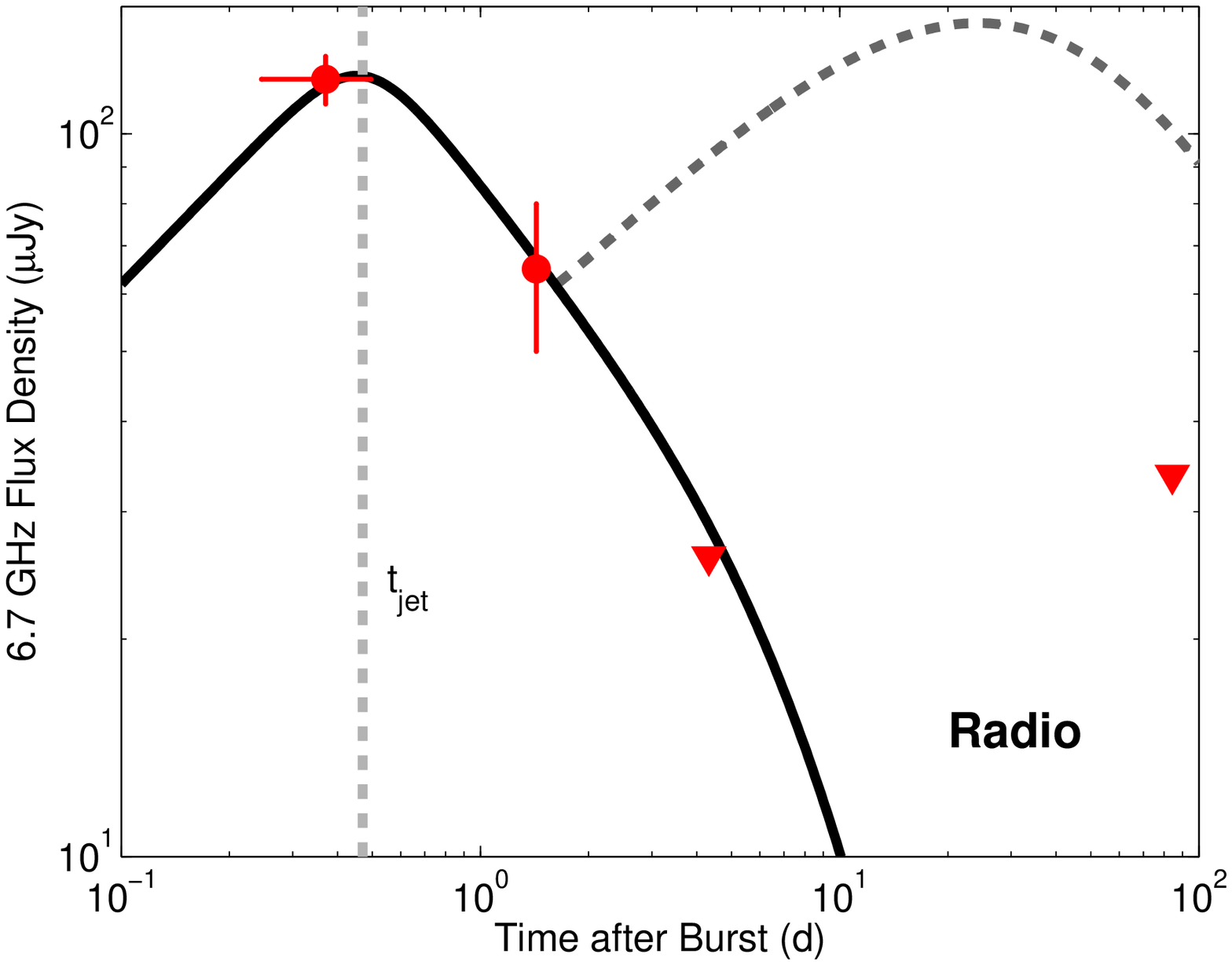}
\includegraphics*[width=0.5\textwidth,clip=]{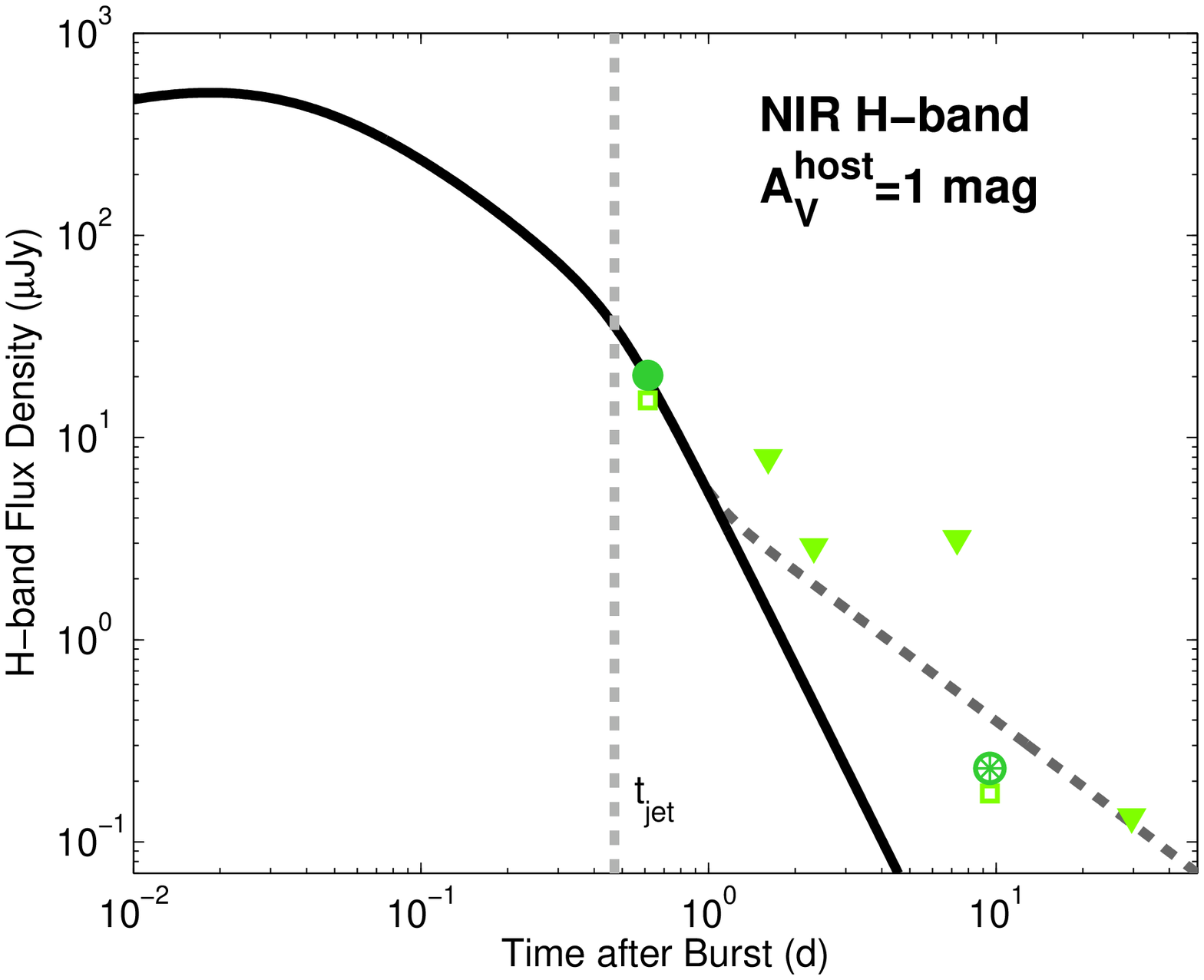} \\
\includegraphics*[width=0.5\textwidth,clip=]{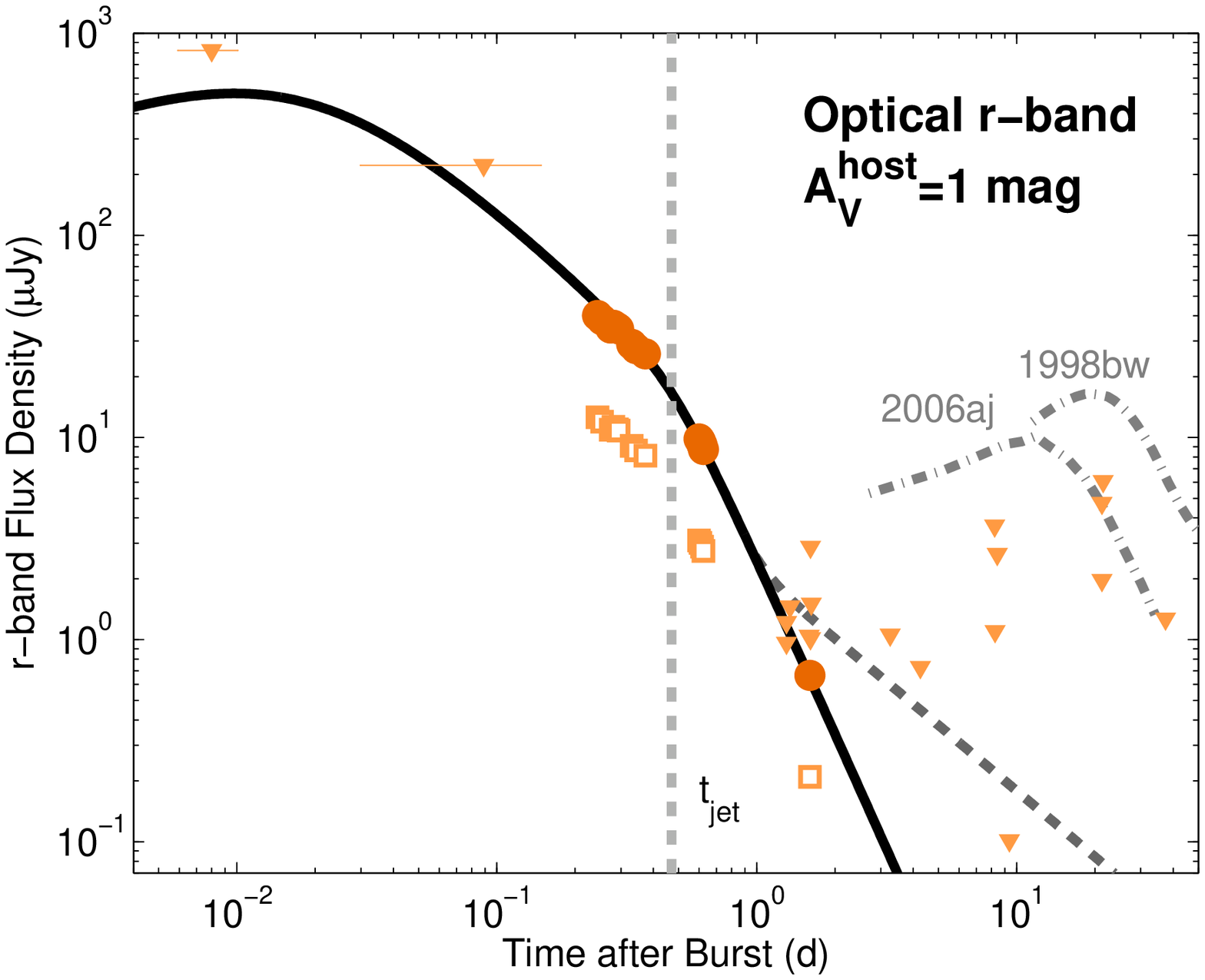} 
\includegraphics*[width=0.5\textwidth,clip=]{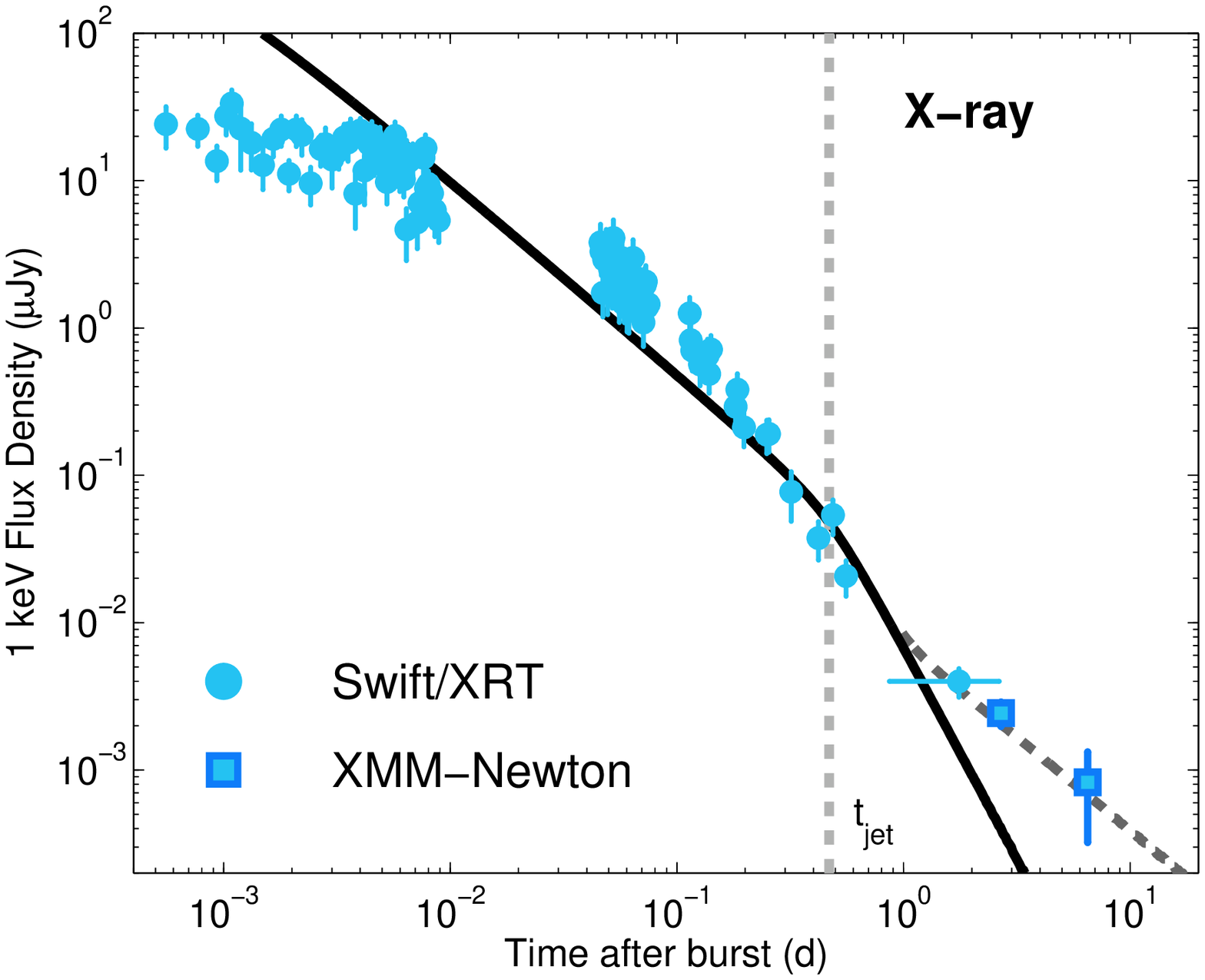} 
\caption{Radio through X-ray afterglow light curves of \grb. Error
bars correspond to $1\sigma$ confidence, and triangles denote
$3\sigma$ upper limits. The afterglow model is shown as a black line,
while the jet break time of $t_j\approx 0.47$~d is marked by a
vertical grey dashed line. Also shown is a model with energy injection
(dark grey dashed line) that fits the X-ray excess but violates the
detections and limits in the optical, NIR, and radio bands. {\it Top
left:} 6.7~GHz observations with the VLA (red). {\it Top right:}
$H$-band observations (green; \citealt{bfc13,dtr+13,tlf+13}), where
$JK$-band observations are extrapolated to $H$-band using $\beta_{\rm
opt}=-2$. The observed values (open green squares) are corrected for
$A_V^{\rm host}=1$~mag (filled green symbols). The circled asterisk at
$\delta t \approx 9$~d is the ``kilonova'' associated with \grb\
\citep{bfc13,tlf+13}. {\it Bottom left:} Optical $r$-band observations
(orange; this work, \citealt{cpp+13,dtr+13,tlf+13}), where $giz$-band
observations are extrapolated to $r$-band using $\beta_{\rm
opt}=-2$. The observed values (open orange squares) are corrected for
$A_V^{\rm host}=1$~mag (filled orange symbols). The displayed upper
limits (orange triangles) are also corrected for extinction. Also
shown are the optical light curves of GRB-SN\,1998bw
\citep{gvv+98,csc+11} and GRB-SN\,2006aj (dot-dashed lines;
\citealt{mha+06}) corrected for extinction and redshifted to
$z=0.3565$. {\it Bottom right:} Observations from \swift/XRT (blue
circles) and \xmm\ (blue squares) at 1~keV.
\label{fig:lc}}
\end{figure*}

To investigate the temporal behavior of the optical/NIR afterglow, we
interpolate all of the available afterglow data (Table~\ref{tab:obs})
to the optical $r$-band using the observed spectral slope
(Section~\ref{sec:sed}) and then correct these fluxes for $A_V^{\rm
host}=1$~mag. The resulting temporal behavior of the optical afterglow flux is
characterized by a broken power law (Figure~\ref{fig:lc}). Invoking a
broken power law model with two segments, we find pre- and post-break
decay indices of $\alpha_1 \approx -1.26 \pm 0.05$ and $\alpha_2 =
-2.73 \pm 0.08$, with a break time of $t_b =
0.47^{+0.02}_{-0.06}$~d. The required change in the temporal index is
therefore $\Delta\alpha \approx 1.5$. Although there are several
possibilities that can explain breaks in GRB afterglow light curves,
most of these cannot explain the large change in slope and the steep
post-break decline seen here. For instance, the transition of the
cooling frequency across the band predicts $\Delta\alpha=0.25$
\citep{spn98}, while the cessation of energy injection observed in
long GRBs typically leads to $\Delta\alpha \approx 0.7$ with
moderate decline rates of $\alpha_{1} \approx -0.5$ and $\alpha_{2}
\approx -1.2$ \citep{nkg+06,zfd+06,lzz07}. A steep drop
in the density is predicted to cause maximum changes of $\Delta\alpha
\approx 0.4$ for density contrasts of $\sim 10$ \citep{ng07}, and
would require a density contrast of $\gtrsim 1000$ to account for
$\Delta\alpha=1.5$.

We therefore conclude that the temporal steepening is instead a jet break,
when the edge of a relativistically-beamed outflow becomes visible to
the observer \citep{sph99,rho99}. In this scenario, the post-break
flux declines as $t^{-p}$ \citep{sph99}. Indeed, we find good
agreement between $\alpha_2 = -2.7 \pm 0.1$ determined from the
optical light curve and $p = 2.55 \pm 0.15$ independently
determined from the broad-band SED (Section~\ref{sec:sed}).

In addition, since the radio band lies between $\nu_a$ and $\nu_m$
(Section~\ref{sec:sed}), the radio flux density should increase as
$F_{\nu} \propto t^{1/2}$ in the spherical regime \citep{gs02}, while
the flux will decrease as $F_{\nu} \propto t^{-1/3}$ in a post-jet
break scenario \citep{sph99}. We find that the observed radio flux of
\grb\ declines with $\alpha \approx -0.45$ between $\delta t \approx
0.4$~d and $1.3$~d, demonstrating that the evolution is not isotropic
(Figure~\ref{fig:lc}). Thus, the temporal behavior of both the
optical and radio afterglows support a jet break at $t_j \approx
0.47$~d, making this the first detection of a jet break in the radio
afterglow of a short GRB.

In conjunction with the energy, density, and redshift, the time
of the break can be converted to a jet opening angle
\citep{sph99,fks+01},

\begin{equation}
\theta_j=9.51 t_{j,{\rm d}}^{3/8}(1+z)^{-3/8}E_{\rm K,iso,52}^{-1/8}n_0^{1/8} \text{ deg}
\label{eqn:jb}
\end{equation}

\noindent where $t_{j,{\rm d}}$ is in days, $E_{\rm K,iso,52}$ is in
units of $10^{52}$~erg and $n_0$ is in units of cm$^{-3}$. For the
ranges of $E_{\rm K,iso}$ and $n_0$ in
Equations~\ref{eqn:constr4}-\ref{eqn:constr5}, we calculate $\theta_j
\approx 4-14^{\circ}$. However, given a NS-NS/NS-BH progenitor as
indicated by the likely detection of a kilonova \citep{bfc13,tlf+13},
the circumburst density is likely more typical of ISM densities ($n
\lesssim 1$~cm$^{-3}$), leading to $\theta_j \approx 4-8^{\circ}$.

\section{Late-time X-ray Excess}
\label{sec:xex}

The synchrotron afterglow model with a jet break at $t_j \approx
0.47$~d provides a good match to the radio and optical light curves
(Figure~\ref{fig:lc}).  However, unlike the optical afterglow
behavior, we do not observe significant steepening in the X-ray light
curve, and instead the afterglow flux at $\delta t \gtrsim 0.03$~d can
be characterized by a single power law with $\alpha_X = -1.88 \pm
0.15$. Thus, our afterglow model underpredicts the X-ray flux by a factor of $\approx
5$ at $\delta t \gtrsim 2$~d (see Figure~\ref{fig:lc}). We note that
\citet{tlf+13} and \citet{dtr+13} previously claimed that the same
broken power-law fits both the optical and X-ray data, but this was
based on the XRT data alone. The \xmm\ observations do not support
these claims. We are thus motivated to consider an additional
energy source for this excess X-ray emission, which
follows $L_{\rm X} \simeq 4\times 10^{43}t_d^{-1.88}$~erg~s$^{-1}$
(where $t_d$ is in units of days).

One possible source of late-time excess X-ray emission is continued
energy injection into the blastwave from ongoing central engine
activity \citep{zm02,zfd+06}. We first consider continuous power law
energy injection with $L \propto t^q$. We find that energy injection
beginning at $\delta t \approx 1$~d characterized by $q=0.3$ (e.g., $E
\propto t^{1.3}$) provides an adequate fit to the late-time X-ray
light curve (Figure~\ref{fig:lc}). Assuming that the injection extends
to $\delta t \approx 10$~d, the energy injection factor is $\approx
9.5$. However, this energy injection model violates both the optical
and radio light curves; in particular, the radio upper limits are a
factor of $\approx 3$ below the predicted flux with energy injection
(Figure~\ref{fig:lc}). Therefore, continuous energy injection is not a
viable explanation for the excess X-ray emission.

We are therefore motivated to consider a source of emission that
predominantly contributes in the X-rays with negligible effects on the
other bands.  We focus on scenarios in which this emission originates
from the central engine, but which we now assume can be viewed
directly through the merger ejecta (we justify this assumption below).
We first consider ongoing accretion (``fall-back'') onto the
newly-formed black hole following a compact object merger.  By
extrapolating the matter trajectories from numerical simulations of
the merger process to late times, \citet{ros07} predicts accretion
luminosities of $L_{\rm acc} \sim 10^{43}-10^{45}$ erg s$^{-1}$ on a
timescale of $\sim 1$ d, with a temporal decay of $L_{\rm acc} \sim
t^{-\alpha}$ similar to the canonical prediction of $\alpha = -5/3$
for a tidally disrupted star \citep{ree88}. Assuming a radiative
efficiency of $\gtrsim 10\%$, this scenario is well matched to the
X-ray light curve of \grb. However, such efficiencies are optimistic
and more detailed models of the fall-back process from compact object
mergers \citep{rb09} predict much lower X-ray luminosities of $L_{\rm
X} \lesssim 10^{-3}L_{\rm acc}$, which would not be large enough to
explain the observed excess.

Another possibility is that the X-ray emission is powered by the
 spin-down of a massive magnetar remnant \citep{mqt+08,bmt+12,zha13},
 a process that has been used to explain putative plateaus in the
 X-ray afterglows of short GRBs \citep{rom+13}. Such remnants may be
 at least temporarily stable to gravitational collapse, if they rotate with spin
 periods of $\lesssim$few ms (e.g.~\citealt{opr+10}; \citealt{gp13}). The remnant may also acquire strong
magnetic fields of $\gtrsim 10^{14}-10^{15}$~G, similar to those of
Galactic magnetars \citep{dt92,zm13}.  The magnetar model predicts
that the late-time spin-down luminosity should decay as $L_{\rm sd} \propto
t^{-2}$, consistent with the observed temporal decay of $\alpha_X =
-1.88 \pm 0.15$.  The predicted spectrum is $F_{\nu} \propto \nu^{-1}$
(\citealt{met13}), also consistent with the observed spectral index
$\beta_X \approx -1.2 \pm 0.1$. Fitting the entire X-ray light curve
with a magnetar model characterized by the duration and luminosity of
the plateau \citep{zm02}, and assuming $M_{\rm NS}=1.4-2.5$~$M_{\odot}$
and $R_{\rm NS}=10^{6}$~cm, we find best-fit parameters of $B \approx
2 \times 10^{16}$~G and $P \approx 15-25$~ms (where higher mass
corresponds to slower spin periods). However, such slow spin periods
are likely unphysical in the merger context due to the substantial
angular momentum of the initial binary. Instead, assuming a more
reasonable initial spin period of 1~ms, the required magnetic field
strength to produce the observed X-ray luminosity at $\gtrsim 1$~d
assuming 10\% radiative efficiency, is $B \approx 10^{15}$~G, but such
a model would under-predict the light curve at $\delta t \lesssim
3000$~s by a factor of a few. We thus conclude that the magnetar
scenario could potentially explain the late-time X-ray excess for
$\delta t \gtrsim 3000$~s.

To justify that we can observe the central engine directly, as is
required in either the fall-back or magnetar models, the merger ejecta
must be transparent to soft X-rays.  Due to the high bound-free X-ray
opacity of neutral matter, this in turn requires that the engine be
sufficiently luminous to re-ionize the merger ejecta
\citep{met13}. The ejecta becomes transparent to X-rays once two
conditions are satisfied: (1) the opacity becomes dominated by
electron scattering ($\kappa_{\rm bf}/\kappa_{\rm es} \lesssim 1$,
where $\kappa_{\rm bf}$ and $\kappa_{\rm es}$ are the bound-free and
electron scattering opacities, respectively) and (2) the electron
scattering optical depth, $\tau_{\rm es}$, through the ejecta
decreases to $\lesssim 1$. Assuming ejecta mainly composed of
hydrogen-like iron\footnote{Outflows along the polar direction arise
chiefly from the accretion disk and are expected to be composed of
Fe-like nuclei (e.g.~\citealt{mqt+08}).} and an ejecta temperature of
$T_{\rm ej}=10^{4}$~K, we derive the following expressions (see
Appendix for details):

\be \frac{\kappa_{\rm bf}}{\kappa_{\rm es}}
\approx 0.13 \left(\frac{L_{X}}{4\times 10^{43}{\rm
erg\,s^{-1}}}\right)^{-1}\left(\frac{M_{\rm
ej}}{10^{-3}M_{\odot}}\right)\left(\frac{t}{\rm 1~
d}\right)^{-1}\left(\frac{v_{\rm ej}}{c}\right)^{-1} ,
\label{eq:ion}
\ee

\be 
\tau_{\rm es} = \rho_{\rm ej}\kappa_{\rm es}R_{\rm ej} \simeq 0.02\left
(\frac{M_{\rm ej}}{10^{-3}M_{\odot}} \right ) \left( \frac{v_{\rm
    ej}}{c}\right)^{-2} \left ( \frac{t}{1{\rm\, d}} \right)^{-2}
\label{eq:tau}
\ee 

\noindent where $M_{\rm ej}, R_{\rm ej} = v_{\rm ej}t$, and $v_{\rm
ej}$ are the (effective) mass, radius, and velocity of the ejecta
along the observer line of sight, respectively. Using fiducial values
for $M_{\rm ej}$ and $v_{\rm ej}$ of the merger \citep{hkk+13,fm13},
we find that both conditions are satisfied on timescales $\delta t
\gtrsim $ few hr given the observed X-ray luminosity.  We thus
conclude that the ejecta are indeed transparent to the soft X-rays at
late times, supporting the idea that direct radiation from the central
engine could produce the observed X-ray excess emission in \grb.

We note that of the $\approx 10$ short GRBs with X-ray observations to
$\delta t \gtrsim 1$~d, two events, GRBs\,050724 and 080503, also
exhibited late-time X-ray excess emission on timescales of $\delta t
\sim 1-2$~d \citep{gbp+06,pmg+09}. However, unlike \grb, these bursts
both had corresponding behavior in the optical bands
\citep{mcd+07,pmg+09}, suggesting that the optical and X-ray emission
were from the same emitting region.

\section{Comparison to Previous Short GRBs}

The broad-band afterglow observations of \grb\ provide the second
detection of a multi-wavelength jet break in a short GRB, and the
first detection of a jet break in the radio band. Radio afterglow
emission has thus far been detected in two short GRBs: GRB\,050724A
\citep{bpc+05}, GRB\,051221A \citep{sbk+06}. The ability to monitor
the radio afterglow of \grb\ at a flux density level of $\lesssim 0.1$
mJy highlights the improved sensitivity of the VLA.

Indeed, the radio evolution can provide an important constraint on the
progenitor. In the context of the compact object binary progenitor,
the radioactive decay of $r$-process elements in the sub-relativistic
merger ejecta is predicted to produce transient emission, termed a
``kilonova'' \citep{lp98,mmd+10,gbj11,rkl+11,rka+13}, which is
expected to peak in the NIR \citep{bk13,kbb13,th13}. Late-time NIR
emission in \grb\ detected with the {\it Hubble Space Telescope}
(Figure~\ref{fig:lc}; \citealt{bfc13,tlf+13}) has been interpreted as
the first detection of a $r$-process kilonova. An alternative scenario
to explain the excess NIR emission of \grb\ may be a wide, mildly
relativistic component of a structured jet \citep{jxf+13} which has
been used to explain the light curve behavior of a handful of long
GRBs (e.g., \citealt{bkp+03,sfw+03,pkg+05,rks+08}). In this scenario,
the predicted radio emission is similarly boosted, and will be
$\approx 80\,\mu$Jy at $\delta t \approx 84$~d, the time of our final
radio observations \citep{jxf+13}. Instead, the non-detection of any
radio emission to $\lesssim 34\,\mu$Jy provides a strong constraint on
the existence of a two component jet, and supports the kilonova
interpretation of the NIR emission.

\begin{figure}
\centering
\includegraphics*[angle=0,width=3.5in]{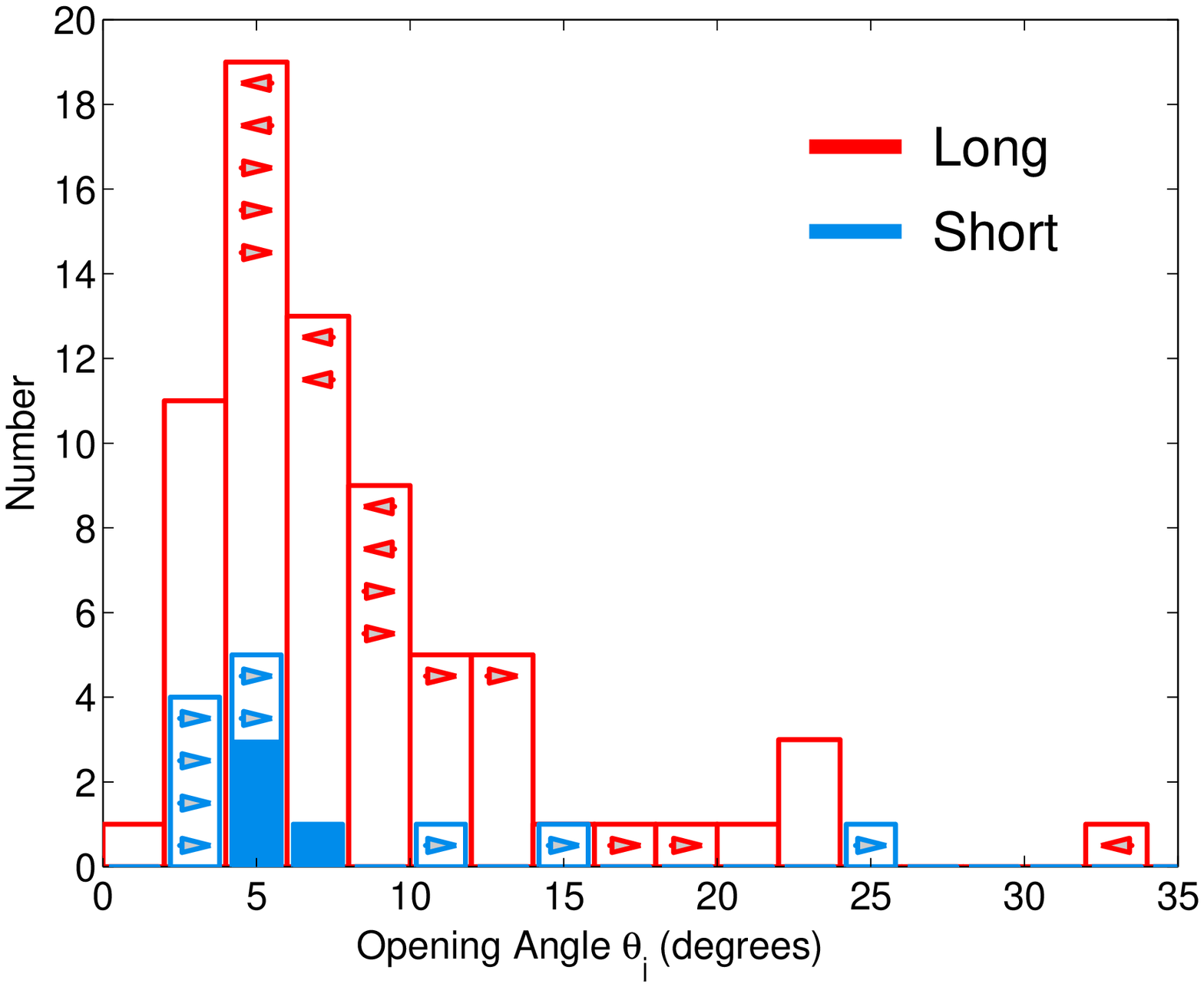}
\caption{Distribution of opening angles for long (red) and short
(blue) GRBs, updated from \citet{fbm+12}. Arrows represent upper and
lower limits. The long GRB population includes pre-\swift\
\citep{fks+01,bkf03,bfk03,ggl04,fb05}, \swift\ \citep{rlb+09,fkg+11},
and {\it Fermi} \citep{cfh+10,gpb+11,cfh+11} bursts. For short GRBs,
the existing measurements are GRB\,051221A ($7^{\circ}$;
\citealt{sbk+06}), GRB\,090426 ($5-7^{\circ}$, assigned $6^{\circ}$
here; \citealt{nkr+11}), GRB\,111020A ($3-8^{\circ}$, assigned
$5.5^{\circ}$ here; \citealt{fbm+12}) and \grb\ ($4-8^{\circ}$,
assigned $6^{\circ}$ here; this work). Short GRB lower limits are from
the non-detection of jet breaks in \swift/XRT data \citep{fbm+12},
{\it Chandra} data for GRBs\, 050724A \citep{gbp+06}, 101219A
\citep{fbc+13}, 111117A \citep{mbf+12,sta+13}, and 120804A
\citep{bzl+13} and optical data for GRBs\,050709 \citep{ffp+05} and
081226A \citep{nkg+12}.
\label{fig:angle}}
\end{figure}

The detection of a jet break in \grb\ leads to an opening angle measurement of
$4-14^{\circ}$, with a more likely range of $4-8^{\circ}$. This
opening angle is the fourth\footnotemark\footnotetext{We note that
\citet{nkg+12} claimed a jet break in the GRB\,090305A afterglow but
this is based on a single optical data point.} such measurement for a
short GRB after GRB\,051221A ($7^{\circ}$; \citealt{sbk+06}),
GRB\,090426 ($5-7^{\circ}$; \citealt{nkr+11}), and GRB\,111020A
($3-8^{\circ}$; \citealt{fbm+12}). From these four short GRB opening
angle measurements, the median is $\langle
\theta_j \rangle \approx 6^{\circ}$ (Figure~\ref{fig:angle}).

The non-detections of jet breaks can provide lower limits on the
opening angles, assuming on-axis orientation, as off-axis observing
angles could disguise jet breaks \citep{vm12,vm13}. Indeed, such
non-detections to timescales of $\sim 1$ day with \swift/XRT have led
to lower limits of $\theta_j \gtrsim 2-6^{\circ}$ \citep{fbm+12},
while monitoring with more sensitive instruments such as \chandra\ and
\xmm\ to timescales of $\sim 1$ week has led to more meaningful limits
of $\theta_j \gtrsim 10-25^{\circ}$ (Figure~\ref{fig:angle};
\citealt{ffp+05,gbp+06,bzl+13}). The search for jet breaks has been
less fruitful in the optical bands, primarily due to the intrinsic
faintness of the optical afterglows and contamination from host
galaxies. Indeed, the sole lower limit from a well-sampled optical
light curve is from GRB\,081226A, with $\theta_j \gtrsim 3^{\circ}$
\citep{nkg+12}, while we conservatively adopt a lower limit of
$\theta_j \gtrsim 15^{\circ}$ for GRB\,050709 based on a
sparsely-sampled optical light curve \citep{ffp+05}. Using the
measured opening angles and lower limits of $\gtrsim 10-25^{\circ}$, a
likely median for short GRBs is $\langle \theta_j \rangle \approx
10^{\circ}$.

The opening angle distribution of short GRBs impacts the true energy
scale and event rate. The true energy is lower than the
isotropic-equivalent value by the beaming factor, $f_b$ ($f_b \equiv
1-{\rm cos}(\theta_j)$, $E=f_bE_{\rm iso}$), while the actual event rate is
increased by $f_b^{-1}$. For \grb, with an opening angle of $\approx
4-8^{\circ}$, the resulting beaming factor is $f_b \approx (0.2-1)
\times 10^{-2}$. Therefore, the true energies are $E_{\gamma} \approx
(0.5-2) \times 10^{49}$~erg and $E_K \approx (0.1-1.6) \times
10^{49}$~erg. The small population of short GRBs with well-constrained
opening angles have beaming-corrected energies of $E_{\gamma} \approx
E_{K} \approx 10^{49}$~erg \citep{sbk+06,bgc+06,gbp+06,fbm+12,nkg+12},
roughly two orders of magnitude below the inferred true energies for
long GRBs \citep{fks+01,bfk03,kb08,rlb+09}.

The true event rate is elevated compared to the observed rate by
$f_b^{-1}$. The current estimated observed short GRB volumetric rate
is $\approx 10$ Gpc$^{-3}$ yr$^{-1}$ \citep{ngf06}. For a median opening
angle of $\approx 10^{\circ}$, the median inverse beaming factor is
$f_b^{-1} \approx 70$, resulting in a true rate of $\approx
700$~Gpc$^{-3}$ yr$^{-1}$. The observed all-sky event rate of $\approx
0.3$~yr$^{-1}$ within 200~Mpc \citep{gp05} then becomes $\approx
20$~yr$^{-1}$. This rate is comparable to estimates for NS-NS merger
detections with Advanced LIGO/VIRGO \citep{aaa+13}.

\section{Conclusions}

We presented broad-band observations of the afterglow of \grb,
uncovering a jet break in the optical and radio light curves at
$\delta t \approx 0.47$~d. This comprehensive data set marks the first
detection of a jet break in the radio band and the third radio
afterglow detection in nearly a decade of follow-up. The inferred
opening angle is $\theta_{j} \approx 4-8^{\circ}$, leading to true
energy releases of $E_{\gamma} \approx E_{K} \approx 10^{49}$~erg.

We observe excess X-ray emission at $\gtrsim$1 day with no
corresponding emission in the other bands. We rule out energy
injection from ongoing activity from the central engine due to the
non-detection of any radio or optical emission on a similar
timescale. We find that fall-back accretion can explain the late-time
excess only if the radiative efficiency is $\gtrsim 10\%$. Finally, we
consider that the emission is due to the spin-down of a massive
magnetar and find that a model characterized by a spin period of
$\approx 1$~ms and magnetic field of $\approx 10^{15}$~G provides a
good fit to the emission at $\delta t \gtrsim 1$~d, but underpredicts
the X-ray emission at $\lesssim 3000$~s. Furthermore, we show that the
merger ejecta are transparent to soft X-rays (also see Appendix),
ensuring that the engine can be viewed in X-rays.

\grb\ highlights the importance of multi-wavelength afterglow
observations, which provide the only route to constraints on the basic
explosion properties of GRBs. In particular, the radio band is
advantageous because unlike the optical, it does not typically suffer
from host galaxy contamination, and can provide an additional
constraint on the circumburst density. Thus, continued
target-of-opportunity efforts at the VLA will provide
invaluable information on the sub-parsec explosion environments. In
addition, the non-detection of late-time radio emission, coupled with
the detection of NIR excess emission, can provide unambiguous support
for the kilonova interpretation for future events, as it has for \grb.

The opening angle determination for \grb\ is the fourth robust jet
break measurement for a short GRB. Using realistic assumptions for the
opening angle distribution, this implies a conservative volumetric event rate
of $\approx 700$~Gpc$^{-3}$~yr$^{-1}$, and an all-sky event rate
of $\approx 20$~yr$^{-1}$ within 200~Mpc, consistent with the
predictions of NS-NS merger detections with Advanced LIGO/VIRGO. However, the opening angle distribution for wider jets
of $\gtrsim 5^{\circ}$ is poorly constrained, and it is necessary to
continue monitoring short GRB afterglows to late times to
characterize this part of the distribution.

\appendix

\section{Re-ionization Model}

The average density of the freely-expanding ejecta decreases with time
 as $\rho_{\rm ej} \simeq M_{\rm ej}/(4\pi/3 R_{\rm ej}^{3})$, where
 $M_{\rm ej}, R_{\rm ej} = v_{\rm ej}t$ is the ejecta radius, and
 $v_{\rm ej}$ is the ejecta velocity.  The ionization state of the
 ejecta is determined by comparing the absorption rate of ionizing
 photons $\mathcal{R}_{\rm ion}$ = $\mathcal{C}n_{\gamma,\nu \gtrsim
 \nu_1}\sigma_{\nu_1}c$ (per ion) to the rate of recombination
 $\mathcal{R}_{\rm rec} = n_e \alpha_{\rm rec}$, where $n_{\gamma,\nu
 > \nu_1}$ = $L_{\rm X}/4\pi h\nu_1 R_{\rm ej}^{2}c$ is the number
 density of ionizing photons; $\mathcal{C}$ is a constant of order
 unity that depends on the spectrum of the ionizing radiation;
 $L_{\nu_1}$ is the specific X-ray luminosity near the ionization
 threshold energy $h\nu_1 \sim 10$ keV; $\sigma_{\nu_1} \simeq 8\times
 10^{-21}$ cm$^{2}$ is the photoionization cross section at $\nu =
 \nu_1$ and $n_e \simeq \rho_{\rm ej}/2m_p$ is the number density of
 electrons in the ejecta; $\alpha_{\rm rec} \approx 2.0\times
 10^{-10}T_4^{-0.8}$ cm$^{3}$ s$^{-1}$ is the [type 2] recombination
 coefficient (e.g.~\citealt{of06}); and $T_4\sim 1 $ is the ejecta
 temperature in units of 10$^{4}$ K. The above results can be combined
 to determine the ratio of the bound-free $\kappa_{\rm bf} = f_{\rm
 n}\sigma_{\nu_1}/26m_p$ and electron scattering $\kappa_{\rm es} =
 0.2$ cm$^{2}$ g$^{-1}$ opacities of the ejecta, where $f_{\rm n}
 \approx \mathcal{R}_{\rm rec}/\mathcal{R}_{\rm ion} \ll 1$ is the
 neutral fraction (set by the balance between ionization and
 recombination rates). The resulting coefficient is in
 Equation~\ref{eq:ion}, from which it is shown that the ejecta is
 sufficiently ionized for the observed X-rays to originate directly
 from the central engine interior to the ejecta.

One might be concerned that an X-ray source of sufficient luminosity
to ionize the ejecta along the observer line of site (perpendicular to
the merger plane) would also be sufficient to ionize matter ejected in
the equator, the radioactive heating of which powers the optical/NIR
kilonova emission.  Indeed, the red colors of the kilonova result from
the high line opacity of the lanthanides, which would vanish were the
ejecta ionized by the central engine.  However, equation
(\ref{eq:ion}) shows that $\kappa_{\rm bf}/\kappa_{\rm es} \propto
L_{X}^{-1}t^{-1}$ is larger for the equatorial ejecta (due to its
larger mass $M_{\rm ej}$ and lower velocity $v_{\rm ej}$), and that
this ratio increases with time $\propto t$ (since $L_{X} \propto
t^{-2}$, approximately).  The fact that $\kappa_{\rm bf}/\kappa_{\rm
es}$ may be sigificantly $\gg 1$ for the equatorial kilonova
ejecta implies that the latter may remain neutral, preserving the
kilonova emission.  However, more detailed calculations, including the
different (and uncertain) recombination rates of the lanthanides, is
necessary to verify this conclusion.

\acknowledgments

We thank Barry Madore for obtaining Magellan observations. We thank
the VLA staff for rapid response capabilities, and especially Joan
Wrobel for scheduling of VLA observations. The Berger GRB group is
supported by the National Science Foundation under AST- 1107973. BAZ
is supported by an NSF Astronomy and Astrophysics Postdoctoral
Fellowship under award AST-1302954. This paper includes data gathered
with the 6.5 meter Magellan Telescopes located at Las Campanas
Observatory, Chile. This work made use of data supplied by the UK
Swift Science Data Centre at the University of Leicester. This work is
based on observations obtained with XMM-Newton, an ESA science mission
with instruments and contributions directly funded by ESA Member
States and the USA (NASA). Observations were obtained with the VLA
(program 13A-046) operated by the National Radio Astronomy
Observatory, a facility of the National Science Foundation operated
under cooperative agreement by Associated Universities, Inc.

{\it Facilities:} Swift (XRT), XMM-Newton (EPIC-PN), Magellan (IMACS), VLA

\end{document}